\documentclass{ws-procs9x6}

\begin{document}

\title{Studies of TMDs at COMPASS}

\author{H. Wollny on behalf of the COMPASS Collaboration}

\address{Physikalisches Institut, Albert--Ludwigs University of Freiburg,\\ 79104 Freiburg, Germany\\$^*$Heiner.Wollny@cern.ch}

\begin{abstract}

Transverse spin and transverse momentum distribution functions of the
constituents of the nucleon are a crucial input for a complete
description of the nucleon. COMPASS measured such for longitudinally and
transversely polarized deuterons and protons. In the following we will
focus on recent results from the 2007 transverse proton data and on the
results for unpolarized deuterons.

\end{abstract}

\keywords{SIDIS, polarized target, Transversity, TMD, Collins, Sivers}
\bodymatter

\section{Introduction}

At leading order and integrating over transverse quark momenta, three
parton distribution functions (PDFs) are needed for a complete
description of the nucleon. The quark density $q(x)$, the helicity
distribution $\Delta q(x)$ and the transversity distribution $\Delta_T
q(x)$. The first PDF, $q(x)$, describes the probability to scatter off a
quark inside the nucleon carrying the momentum fraction $x$. The
helicity PDF, $\Delta q(x)$, is defined for a longitudinal polarized
proton as the difference of the probabilities that the struck quark
carries momentum fraction $x$ and its spin is parallel or anti-parallel
to the spin of the parent nucleon. The transversity PDF, $\Delta_T
q(x)$, is analogously defined to $\Delta q(x)$, however for a
transversely polarized nucleon. Because $\Delta_T q(x)$ is chiral-odd it
cannot be measured in inclusive DIS. However it can be accessed in
semi-inclusive deep-inelastic scattering in combination with a
chiral-odd fragmentation function. Three different channels to address
the transversity distribution, have been analysed, both for deuteron and
proton targets. They will be discussed in Sec.~\ref{sec:transversity}.

If one considers intrinsic transverse momenta of the quarks inside the
nucleon, several further PDFs, so called transverse momentum dependent
distribution functions (TMDs), contribute to the SIDIS
cross-section, which will be discussed in Sec.~\ref{sec:tmds}. 

\section{The COMPASS Experiment}

COMPASS is a fixed target experiment situated at the M2 beam line of the
SPS accelerator at CERN. It has a wide physics program dedicated to the
study of the nucleon spin structure and of hadron spectroscopy. In the
years 2002, 2003, 2004 and 2006 data with a polarized $^6$LiD target
were taken. In the years 2002-2004 approximately 20\,\% of the data
taking was dedicated to transverse target polarization. In the year 2007
COMPASS took data with a NH$_3$ target, equally shared between
longitudinal and transverse target polarization. After the years 2008
and 2009, dedicated to baryon and meson spectroscopy measurements with
hadron beams, in 2010 COMPASS continues with muon beam and transversely
polarized protons.\\ The target consists of several cells along the beam
direction, which are oppositely polarized. The polarization is
periodically reversed to reduce systematic effects due to the different
acceptances of the cells.  The detector consists of two open field dipole
magnets, allowing the detection of particles with $\pm
180$\,mrad. Particle identification is done with a Ring Imaging Cerenkov
detector, two hadron calorimeters and muon filters. For a detailed
description we refer to reference~\cite{Abbon:2007pq}.

\section{Transversity}
\label{sec:transversity}

{\bf Collins Asymmetry:}
\label{sec:Collins}
The chiral-odd transversity distribution can be measured in SIDIS
involving the chiral-odd Collins fragmentation function $\Delta_T^0
D_q^h$~\cite{Collins:1992kk}, which describes the fragmentation of
transversely polarized quarks into unpolarized hadrons. According to
Collins the fragmentation leads to an asymmetry $A_{Coll}$ of the number
of produced hadrons in $\sin(\phi_h+\phi_S-\pi)$. Here $\phi_h$ is the
azimuthal angle, measured around the direction of the virtual photon, of
the hadron with respect to the scattering plane and $\phi_S$ is the
azimuthal angle between the spin of the initial quark and the scattering
plane~\cite{Alexakhin:2005iw,Ageev:2006da,:2008dn}. In first order the
measured asymmetry $A_{Coll}$ is proportional to a convolution over
intrinsic quark transverse momenta of the transversity distribution and
the Collins fragmentation function measured at
Belle~\cite{Seidl:2008xc}. Because of the convolution, assumptions about
the transverse momentum dependence of the distribution and the
fragmentation functions have to be made to extract
transversity~\cite{Anselmino:2008jk}. In addition the $Q^2$ evolution of
the Collins fragmentation function from Belle to COMPASS energies is not
known, introducing a further uncertainty.\\
{\bf Dihadron Interference:}
\label{sec:Dihadron}
Another possibility to access the transversity distribution is to
measure it in SIDIS in combination with the polarized dihadron
interference fragmentation
function (DiFF)~\cite{Efremov:1992pe,Artru:1995zu,Radici:2001na,Bacchetta:2003vn}. Here,
the fragmentation of a transversely polarized quark into two unpolarized
hadrons leads to an azimuthal modulation in $\sin( \phi_R + \phi_S
-\pi)$ in the number of produced hadron pairs. Here $\phi_S$ is defined
as above and $\phi_R$ is the azimuthal angle, measured around the
direction of the virtual photon, between the vector \mbox{$\vec{R} =
(z_2 \vec{P}_1 - z_1 \vec{P}_2)/(z_1 + z_2)$} and the scattering
plane. Where $P_1$ and $P_2$ are the momenta of the two hadrons and
$z_1$ and $z_2$ are their fractional energies. For oppositely charged
hadron pairs $P_1$ is per definition the momentum of the hadron with positive charge,
otherwise $P_1$ is the momentum of the hadron with the largest energy
fraction. Here, the measured asymmetry $A_{RS}$ is proportional to a
product of the transversity distribution and the polarized DiFF, recently measured at
Belle~\cite{Vossen:2009xz}. Therefore, no assumptions about transverse
momentum dependences of distribution and fragmentation functions are
needed, to extract transversity. In addition, the $Q^2$ evolution of the
polarized DiFF from Belle energies to COMPASS
energies is known~\cite{Ceccopieri:2007ip}. Hence, the extraction of
transversity with this channel is from theoretical point of view
cleaner than the extraction from the Collins asymmetries.\\
{\bf Lambda Polarization:}
\label{sec:Lambda}
A third channel to measure transversity is to study $\Lambda$ and $\bar
\Lambda$ polarization in SIDIS with a transversely polarized
target~\cite{Artru:1990wq}. The polarization $P_T^\Lambda$ of the
produced $\Lambda/\bar \Lambda$-hyperons, measured via the parity
violating decay into $\Lambda \rightarrow p\pi^-$ and $\bar \Lambda
\rightarrow \bar p\pi^+$, respectively, is used as a polarimeter of the
initial transverse spin of the fragmenting quark. It is in first order
proportional to a product of transversity and the fragmentation function
$\Delta_T D_q^\Lambda$, which describes the fragmentation of a
transversely polarized quark into a transversely polarized
$\Lambda$-hyperon.

\section{Transverse Momentum Dependent Distribution Functions, TMDs}
\label{sec:tmds}
Taking into account intrinsic transverse momenta of the quarks in total
eight TMDs are needed for a complete description at
leading-twist~\cite{Mulders:1995dh}.  For unpolarized nucleons the Cahn
effect, the Boer-Mulders TMD and perturbative QCD effects, like gluon
radiation, contribute to $\cos \phi_h$ and $\cos 2\phi_h$ modulations in
the SIDIS cross section of single
hadrons~\cite{Cahn:1978se,Boer:1997nt,Georgi:1977tv}.  For a
transversely polarized nucleon the Sivers TMD $\Delta_0^T
q$~\cite{Sivers:1989cc} is related to the quark angular orbital momentum
inside a transversely polarized nucleon and is therefore of special
interest since this could be a crucial piece to solve the nucleon spin
puzzle. The Sivers effect leads to an azimuthal modulation of the number
of produced hadrons in $\sin(\phi_h - \phi_S)$, where $\phi_h$ and
$\phi_S$ are defined as for the Collins asymmetry discussed in
Sec.~\ref{sec:Collins}. The asymmetry $A_{Siv}$ is proportional to a
convolution over intrinsic quark transverse momenta of the Sivers
function and the well-known unpolarized fragmentation function $D_q^h$.

Recent results of the Sivers function measured with protons and results
of $\cos \phi_h$ and $\cos 2\phi_h$ modulations for unpolarized
deuterons will be reviewed in this article. For the six remaining
azimuthal asymmetries measured with transversely polarized deuterons the
reader is referred to~\cite{Kotzinian:2007uv}. Detailed results of
azimuthal asymmetries measured with longitudinally polarized deuterons
can be found in~\cite{Savin:2010du}.

\section{Event selection}

Kinematic cuts on the squared four momentum transfer $Q^2 >
1$\,(GeV/$c$)$^2$, on the fractional energy transfer of the muon $0.1 <
y < 0.9$ and the hadronic invariant mass $W > 5$\,GeV/$c^2$ are applied
to select DIS events. The selection of charged hadrons differs slightly
for the various analyses. However, for the single hadron analyses in
general a fractional energy $z > 0.2$ of the observed final hadron is
required to select hadrons in the current fragmentation region. In
addition a transverse momentum $p_T^h> 0.1$\,GeV/$c$ of the hadron with
respect to the virtual photon is required to ensure a good definition of
$\phi_h$. Muons are rejected in demanding a minimal energy deposit in
the hadronic calorimeters.\\ For the hadron pair analysis $z_{1,2} >
0.1$ is required for each hadron and for the sum $z_1+z_2 < 0.9$ to
reject exclusively produced $\rho^0$-mesons. In order to have a good
definition of $\phi_R$ a cut on $R_T > 70$\,MeV/$c$, the transverse
component of $\vec R$ with respect to the virtual photon direction, is
applied.

\section{Results}

\subsection{Transversity}

\begin{figure}
    \begin{center}
\psfig{file=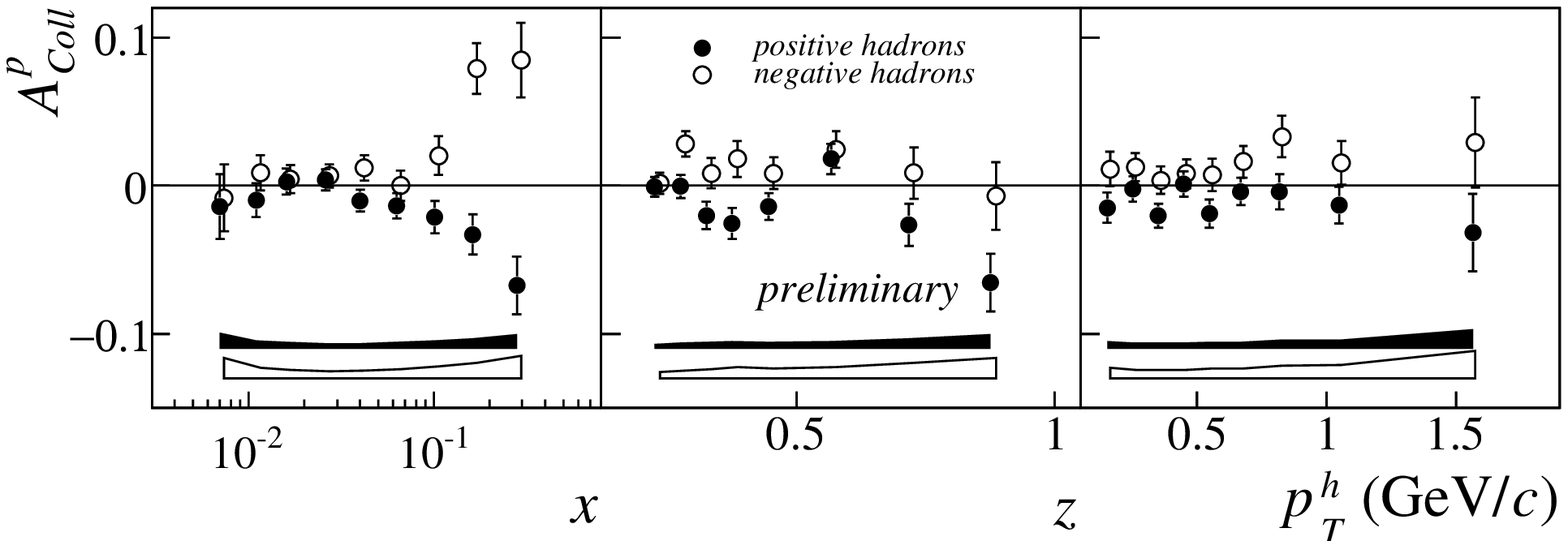,width=0.9\textwidth,trim= 20 5 0 0,clip}
\caption{Collins asymmetries for positive and negative hadrons as a function of $x$, $z$ and $p_T$. The horizontal bands indicate the systematic uncertainties.}
\label{pic:collins}
    \end{center}
    \begin{center}
\psfig{file=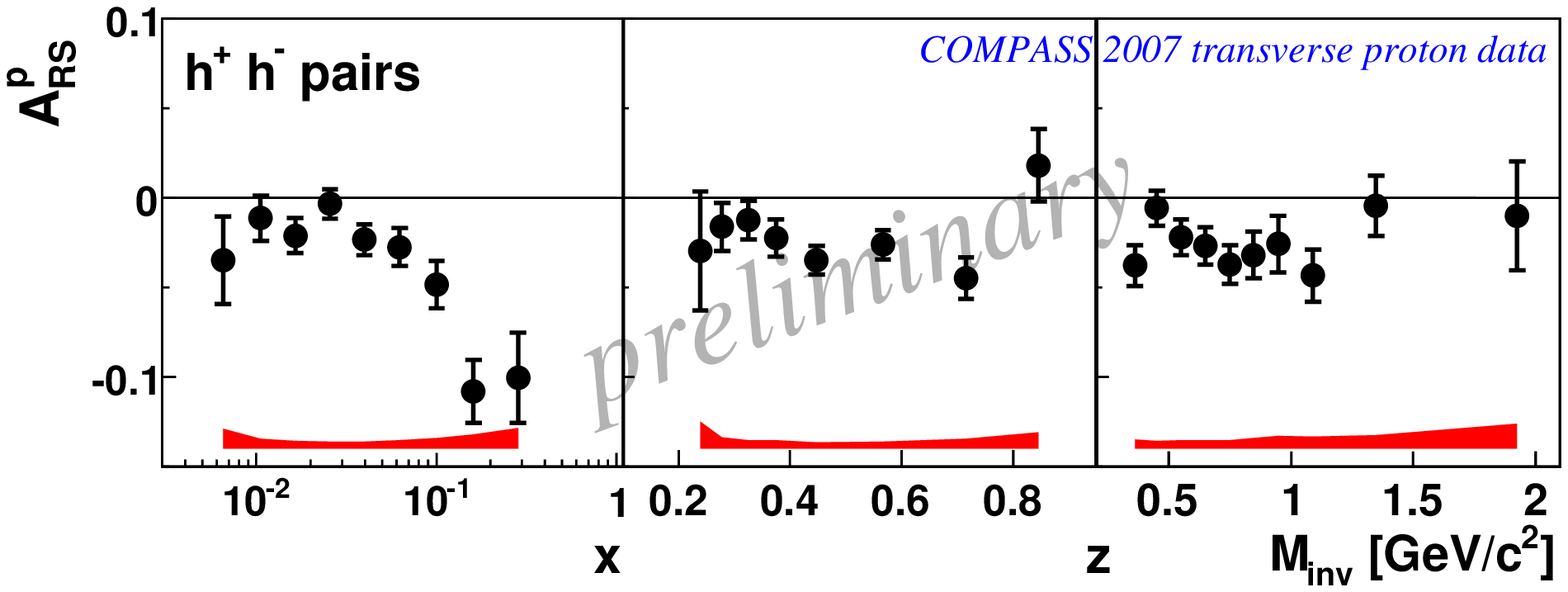,width=0.9\textwidth,trim= 10 0 10 0,clip}
\caption{Dihadron asymmetries $A_{RS}^p$ for oppositely charged hadron pairs as a function of $x$, $z$ and $M_{inv}$. The horizontal band indicates the systematic uncertainties.}
\label{pic:dihadrons}
\vspace{-3mm}
    \end{center}
\end{figure}

The Collins asymmetries $A^p_{Coll}$ have been evaluated for charged
hadrons in bins of Bjorken $x$, energy fraction $z$ and transverse
momentum $p_T^h$. The results for the full 2007 proton statistics~\cite{Fischer:2010} are
shown in Fig.~\ref{pic:collins}. For positive hadrons the asymmetry is
negative and for negative hadrons it is positive. For both charges the
size of the asymmetry increases with $x$ and is compatible in their
strengths. The values agree both in magnitude and in sign with the
previous measurements of HERMES~\cite{:2010ds}, which have
been performed at a considerably lower electron beam momentum of 27.5\,GeV/$c$. From the
transversely polarized deuteron data Collins asymmetries for unidentified charged 
hadrons, charged pions and charged kaons have been
extracted~\cite{Alexakhin:2005iw,Ageev:2006da,:2008dn}. All asymmetries
found to be small and compatible with zero within the statistical
errors. Hence for deuteron the transversity distribution must be small
or even vanishing due to isospin symmetry, because the proton results
and the Belle results~\cite{Seidl:2008xc} confirmed the existence of a non-zero Collins
fragmentation function.
The results for the transverse 2007 proton data of the dihadron
asymmetries $A^p_{RS}$ for oppositely charged hadron
pairs~\cite{Wollny:2009eq} are shown as a function of $x$, $z$ and
$M_{inv}$ in Fig.~\ref{pic:dihadrons}. A strong asymmetry is observed in
the valence $x$-region. In the invariant mass the asymmetry is negative
over the whole range. Compared to the Collins asymmetries, the size of
$A^p_{RS}$ is approximately a factor of two larger, emphasizing the good
analyzing power of the dihadron interference fragmentation function to
measure transversity.
From the transversely polarized deuteron data
COMPASS extracted dihadron asymmetries for unidentified $h^+h^-$-pairs
and identified $\pi^+\pi^-$, $K^+K^-$, $\pi^+K^-$ and
$K^+\pi^-$-pairs~\cite{Joosten:2005vu,Joosten:2007zz,Vossen:2007mh}. In
addition also different charge combinations have been analysed,
considering the two most energetic hadrons in each event. All
asymmetries found to be small and compatible with zero within the
statistical errors.\\
The polarization of $\Lambda$ and $\overline{\Lambda}$-hyperons measured
with transversely polarized $^6$LiD and NH$_3$ targets, respectively,
have been evaluated in bins of $x$ and
$z$~\cite{Ferrero:2007zz,Kang:2010}. For both targets the evaluated
polarizations of $\Lambda$ and $\overline{\Lambda}$-hyperons are small
and compatible with zero within the errors. This indicates, taking into account
the sizeable Collins and dihadron asymmetries measured for protons,
that the analyzing power of this channel seems to be small.

\subsection{Sivers-Function}
\begin{figure}
    \begin{center}
\psfig{file=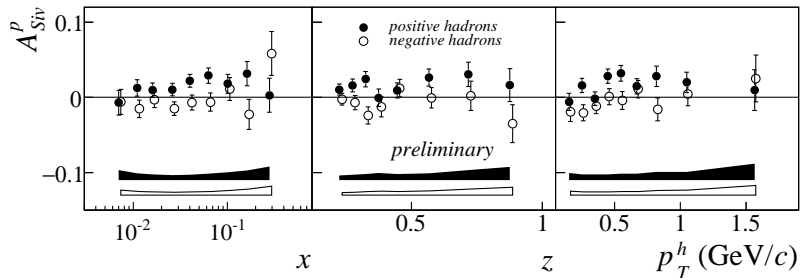,width=0.91\textwidth,trim= 20 5 0 0,clip}
\caption{Sivers asymmetries for positive and negative hadrons as a function of $x$, $z$ and $p_T$. The horizontal bands indicate the systematic uncertainties.}
\label{pic:sivers}
 \end{center}
\end{figure}

The Sivers asymmetries for charged hadrons from the transverse 2007
proton data~\cite{Fischer:2010} are shown, as a function of $x$, $z$ and
$p_T^h$, in Fig.~\ref{pic:sivers}. In addition to the indicated
systematic uncertainty, for positive hadrons an additional systematic
uncertainty of $\pm$0.01 is assigned reflecting a time variation of the
spectrometer acceptance between the first and the second half of data
taking used for this analysis. The asymmetry for positive hadrons is
significantly positive over almost the complete $x$-range.  The
asymmetry for negative hadrons is small and compatible with zero within
the statistical errors. Both results show the same trend as the previous
measurement of HERMES~\cite{:2009ti}. However, the magnitude of the
asymmetry for positive hadrons is smaller than the one of HERMES. This
might be explained by a dependence on $W$, the invariant mass of the
photon-nucleon system. Studies indicate that the asymmetry seems to be
only sizeable at small $W$, where HERMES measures and approach zero at
high $W$. The Sivers asymmetries for unidentified charged hadrons,
charged pions and kaons extracted from transversely polarized
$^6$LiD~\cite{Alexakhin:2005iw,Ageev:2006da,:2008dn} were all found to
be small and compatible with zero within the statistical errors.

\subsection{Azimuthal Asymmetries from Unpolarized Deuterons}

\begin{figure}
    \begin{center}
\psfig{file=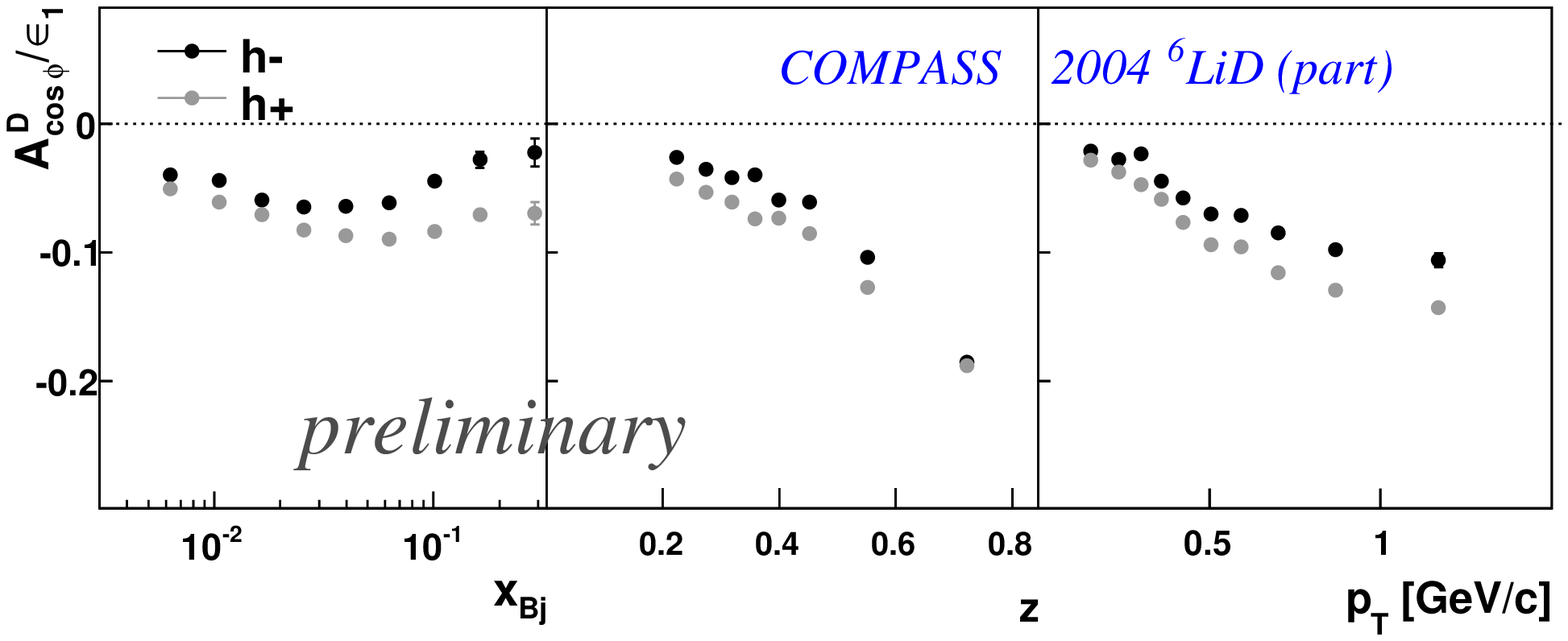,width=0.88\textwidth,trim= 10 47 0 5,clip}\\
\psfig{file=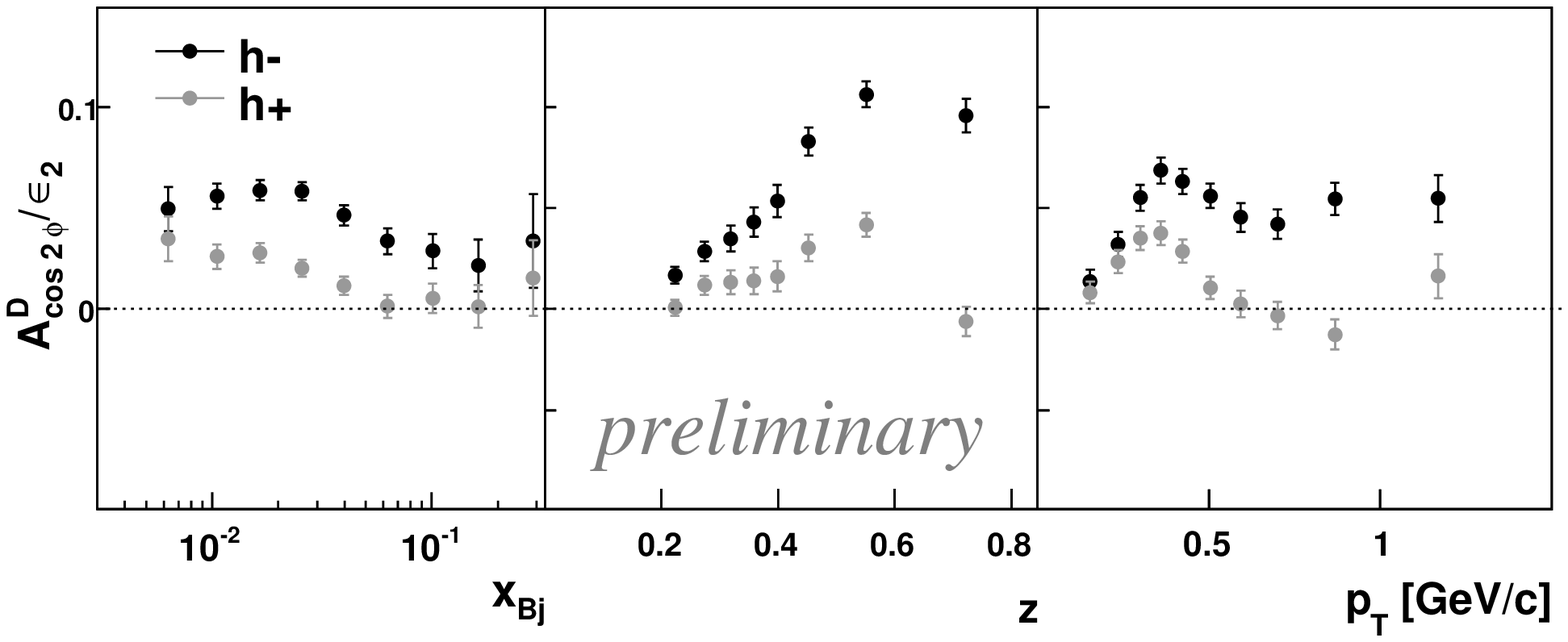,width=0.88\textwidth,trim= 10 5 0 12,clip}
\caption{Unpolarized asymmetries for charged hadrons as a function of $x$, $z$ and $p_T$. The kinematical factors are: $\epsilon_1 = (2-y)\sqrt{1-y}/(1+(1-y)^2)$ and $\epsilon_2 = (1-y)/(1+(1-y)^2)$.}
\label{pic:unpol}
\vspace{-5mm}
    \end{center}
\end{figure}

An unpolarized deuteron sample is obtained by combining data samples
with opposite target polarization taken in the year 2004. The magnitudes
of $\cos \phi_h$ and $\cos 2\phi_h$ modulations for charged
hadrons~\cite{Kafer:2008ud} as a function of $x$, $z$ and $p_T$ are
shown in Fig.~\ref{pic:unpol}. Large asymmetries up to 20\,\% are found
for the $\cos \phi_h$ modulation and asymmetries in the order of 5\,\%
for the $\cos 2\phi_h$ modulation. The trend of the asymmetries for
positive and negative hadrons are similar, however, the magnitudes of
the two differs significantly. The systematic uncertainties have been
evaluated to be sizeable. The dominant contribution is given by the
acceptance corrections determined with Monte Carlo simulations.

\section{Summary and Outlook}

Recent results of COMPASS measurements related to transversity and
transverse momentum dependent distribution functions have been
presented. In the year 2010 and 2011 COMPASS will continue its
measurements with transversely and longitudinally polarized protons,
which will significantly reduce the statistical errors.

\bibliographystyle{ws-procs9x6}
\bibliography{proceeding_wollny_jlab2010.bib}

\end{document}